\documentclass[pra,aps,showpacs,amsmath,amssymb,twocolumn]{revtex4}

\usepackage{graphicx}
\usepackage{dcolumn}
\usepackage{bbm}

\newcommand{\be}{\begin{equation}}
\newcommand{\ee}{\end{equation}}
\newcommand{\ba}{\begin{array}}
\newcommand{\ea}{\end{array}}
\newcommand{\bqa}{\begin{eqnarray}}
\newcommand{\eqa}{\end{eqnarray}}
\newcommand{\sv}{{\cal S}} 

\newcommand{\nb}{\bar n}                                        
\newcommand{\bra}[1]{\ensuremath{\langle #1 |}}
\newcommand{\ket}[1]{\ensuremath{| #1 \rangle}}

\newcommand{\ovl}[2]{\ensuremath{\langle #1 | #2 \rangle}}
\newcommand{\matel}[3]{\ensuremath{\langle #1 | #2 | #3 \rangle}}

\begin{document}

\title{Entanglement dynamics under decoherence: from qubits to qudits}

\author{Andr\'e R. R. Carvalho$^{1}$}
\author{Florian Mintert$^{1,2}$}
\author{Stefan Palzer$^{1}$}\altaffiliation{Permanent address: Fakult\"at f\"ur Physik/FMF, Stefan-Meier-Str. 21, D-79104 Freiburg}
\author{Andreas Buchleitner$^1$} 
\affiliation{$^1$Max-Planck-Institut f\"ur Physik komplexer Systeme,
  N\"othnitzer Strasse 38, D-01187 Dresden\\
$^2$ Department of Physics, Harvard University, Cambridge, MA 02138\\}

\date{\today}

\begin{abstract}
We investigate the time evolution of entanglement for bipartite systems of
arbitrary dimensions under the influence of decoherence. For qubits, we
determine the precise entanglement decay rates under different
system-environment couplings, including finite temperature effects. For qudits, we show how
to obtain upper bounds for the decay rates and also present exact solutions
for various classes of states.
\end{abstract}

\pacs{03.67.-a,03.67.Mn,03.65.Yz,03.65.Ud}

\maketitle

\section{Introduction}

The production of entangled states and the control of their time evolution
became a major issue in current research in view of the development of quantum
information theory, and all possible applications associated with it. Besides
the formidable experimental advances in this direction, there remains a main
obstacle which is the fragility of entanglement under the unavoidable
interaction with the environment. This coupling of the quantum system with its
surroundings, and the consequent decay of entanglement, motivates
important questions such as to understand its sources, to identify the
characteristic timescales, and, possibly, to find ways to circumvent it. 

To devise appropriate strategies for controlling entangled states under the
effect of environment interaction, the first step is to acquire a deeper
understanding of the dynamics of the decoherence processes themselves. Despite the
rapidly increasing experimental interest in this subject, due to the
possibility of monitoring entanglement dynamics~\cite{roosPRL04}, most of the
theoretical work focused on characterizing static properties of entanglement
for quantum states~\cite{ben96d,wot96,vidal00,vidal02a}. 

Only very recently the question of entanglement decay under environment-induced mixing has been addressed, for some specific states
and environment models, and restricted to the case of two
qubits~\cite{yu_02,yu_03,eberly_04}, probably due to the
lack of a genuine {\it and} computable entanglement measure for systems larger
than that. Nonetheless, new techniques for the derivation of
bounds~\cite{flolb,floqp} of concurrence, one possible
entanglement measure, recently allowed 
a systematic study of entanglement dynamics for more general states,
including multipartite~\cite{arrcmintert} and multi-level systems.

These higher dimensional systems are of great interest since they can enlarge the perspectives of efficient applications in quantum information
 and can also be used to test fundamental aspects of quantum theory. In fact,
 entangled states of two $d$-dimensional quantum systems, the $qudits$, can
 improve measurement resolution~\cite{mitchell_04,boto_00} and are known to
 violate local realism more strongly than
 qubits~\cite{kasz_00,collins_02}. Moreover, they can be used for quantum
 computation~\cite{knill_01,bartlett_02,klimov_03,mfs_05} and also in quantum
 cryptography protocols~\cite{hbp_00,boure_01}, which are
 safer~\cite{cerf_02,durt_03} than their qubit counterparts. Despite the
 importance of such entangled qudits, reflected in the intense activity on
 their production and manipulation in different experimental
 setups~\cite{mair_01,linares_01,howell_02,vaziri_02,thew_04,eisen_04,neves_05},
 their dynamics under the influence of environment interaction
 remained unexplored until now. 

The paper is organized as follows. In Section~\ref{sec_conc} we
will briefly recall a recently developed approach for calculating concurrence
which allows us to investigate entanglement between systems of arbitrary
dimensions. In Section~\ref{sec_env} we will present the different models which
describe the interaction of the system with the
environment. Section~\ref{sec_dyn} is devoted to the analysis of the
entanglement decay rates under decoherence processes, starting from the case
of bipartite qubits. With the available analytical tools, some, still
unknown, features of the decoherence dynamics are presented. The
section closes with the analysis of bipartite qudits. A summary of the
main results of the paper is presented in the concluding Section~\ref{sec_concl}.

\section{Entanglement Measure: Concurrence \label{sec_conc}}

In order to follow the environment-induced time evolution of entanglement, one
needs a measure which satisfactorily deals with mixed states. A commonly used
measure in the context of two qubits is the concurrence~\cite{wot98}, defined
for pure states as
\be
c(\Psi)=\Bigl|\matel{\Psi^\ast}{\sigma_y\otimes\sigma_y}{\Psi}\Bigr|\ ,
\label{wotcon}
\ee
where $*$ stands for complex conjugation performed in the standard, computational basis.
For mixed states it can be formulated as
\be
\label{mixwotcon}
c(\rho)=\inf_{\{p_i,\Psi_i\}}\sum_i\ p_i\ c(\Psi_i)\ ,
\ee
with 
\be
\label{decomp}
p_i>0\ ,\hspace{.2cm} \mbox{and} \hspace{.2cm}
\rho=\sum_ip_i\ket{\Psi_i}\bra{\Psi_i}\ .
\ee

In contrast to most other measures, Eq.~(\ref{mixwotcon}) can be solved algebraically with the well known solution
$c(\rho)={\rm{max}}\left\{\lambda_1-\lambda_2-\lambda_3-\lambda_4,0\right\}$,
in terms of the square roots, $\lambda_i$, of the decreasingly ordered eigenvalues of the matrix
$\rho(\sigma_y \otimes \sigma_y) \rho^* (\sigma_y \otimes \sigma_y)$.

For higher dimensional systems, we will use the geneneralization of concurrence given in~\cite{run01} that 
coincides - though not obviously - with the original 
concurrence~\cite{wot98} if restricted to two-level systems.
For practical purposes, it is convenient to express
concurrence~\cite{flomultipart,floPR05} in terms of an
operator $A$ acting on the product space ${\cal H}\otimes{\cal H}$ of two copies of the system, as
\be
c(\Psi)=
\sqrt{\bigl.\bra\Psi\otimes\bra\Psi\ A\
\ket\Psi\otimes\ket\Psi\bigr.}\,.
\label{concurrence-A}
\ee
The operator 
\be
A=\sum_{\alpha} \ket{\chi_{\alpha}}\bra{\chi_{\alpha}}
\ee
is the projector onto the space spanned by the states
$\ket{\chi_{\alpha}}$ that are anti-symmetric with respect to the
exchange of the copies of either ${\cal H}_1$ or ${\cal H}_2$, {\it i.e.}
\be
\ket{\chi_{\alpha}}=\left(\ket{i_k\,i_l}-\ket{i_l\,i_k}\right)\otimes\left(\ket{j_m\,j_n}-\ket{j_n\,j_m}\right)
\label{chi}
\,.
\ee
The states $\{\ket{i_k}\}$ and $\{\ket{j_m}\}$ form,
respectively, arbitrary local bases of ${\cal H}_1$ and  ${\cal H}_2$,
and $\alpha$ is a label for the multi-index $[k,l,m,n]$.

To extend this construction for the case of mixed states, one should
substitute Eq.~(\ref{concurrence-A}) in the convex-roof,
Eq.~(\ref{mixwotcon}), which can be written, in terms of subnormalized states
$\ket{\psi_i}=\sqrt{p_i}\ket{\Psi_i}$, as  
\be
\label{mixcon}
c(\rho)=\inf_{\{\psi_i\}}\sum_i {\sqrt{\bigl.\bra{\psi_i}\otimes\bra{\psi_i}\ A\
\ket{\psi_i}\otimes\ket{\psi_i}\bigr.}}\,.
\ee
Unfortunately, apart from the two-level case where the operator $A$ can
be written in terms of a single vector
$\ket{\chi}=\left(\ket{01}-\ket{10}\right)\otimes\left(\ket{01}-\ket{10}\right)$,
no exact solutions for Eqs.~(\ref{mixwotcon}) and (\ref{mixcon}) are known
and, hence, one has to rely on numerical
efforts to calculate the concurrence.
Also note that numerical solutions of this
optimization procedure define only an upper
bound for concurrence, since there is no a priori information
available on whether
the global minimum or just a local one has been reached. 

To circumvent this problem, we will use suitable approximations which
provide lower bounds of concurrence~\cite{flolb,floqp} and not only
allow for an efficient numerical approach, but also, in some cases, for exact algebraic
solutions. Starting from a decomposition of the density matrix in terms of
pure (subnormalized) states $\rho=\sum_i \ket{\phi_i}\bra{\phi_i}$
and from the vectors $\ket{\chi_{\alpha}}$, one can define a set of
matrices $T^{\alpha}$, given by~\footnote{Note that $\ket{\chi_{\alpha}} \in {\cal H}_1\otimes {\cal H}_1\otimes {\cal H}_2 \otimes {\cal H}_2$, whereas $\ket{\phi_j}\otimes\ket{\phi_k}\in
{\cal H}_1\otimes {\cal H}_2 \otimes {\cal H}_1 \otimes {\cal
H}_2$. Nevertheless, the two spaces are isomorphic, and it is
straightforward to identify the correspondence between their elements.}
\be
\label{talpha}
T^{\alpha}_{jk} = \ovl{\chi_{\alpha}}{\phi_j}\otimes\ket{\phi_k}\,.
\ee
These are connected to the previously defined operator $A$ through the tensor 
\be
\label{tensor_A}
{\cal A}^{l m}_{j k}=\sum_{\alpha}(T^{\alpha}_{lm})^*T^{\alpha}_{jk}= \bra{\psi_l}\otimes\bra{\psi_m}\ A\
\ket{\psi_j}\otimes\ket{\psi_k}.
\ee
It was shown in~\cite{flolb} that concurrence is bounded from below by
\be
\label{lbound}
c(\rho)\ge {\rm max} \left\{\sv_1-\sum_{i>1} \sv_i,0 \right\},
\ee
where the $\sv_i$ are the decreasingly ordered singular values of 
\begin{equation}
{\cal T}=\sum_\alpha Z_\alpha T^\alpha,\
{\rm with}
{\sum_\alpha|Z_\alpha|^2=1}\ ,
\label{freeparbound}
\end{equation}
that still depend on the choice of the complex parameters $Z_\alpha$.
Although any choice provides a lower bound, one might still wish to
carry out an optimization over $Z_{\alpha}$, though now on a much smaller parameter space and with simpler constraints
than in Eq.~(\ref{mixcon}).
Moreover, also
each matrix $T^{\alpha}$ already provides a lower bound, which can be calculated algebraically.

Finally, let us describe an experimentally motivated
approach to calculate lower bounds of concurrence, the quasi-pure
approximation~\cite{floqp}. Although environmental influences cannot be
avoided completely, under typical experimental conditions one deals with
states which are, at least initially, {\it quasi-pure}. This is, they have
a single eigenvalue $\mu_1$ which is much larger than all the others, and an
approximation based on this condition can be developed. Indeed, using
the spectral decomposition $\rho=\sum_i \mu_i \ket{\Psi_i}\bra{\Psi_i}$
of the density matrix and the previously defined subnormalized states,
one can see that the elements of $\cal A$ defined in
Eq.~(\ref{tensor_A}) are proportional to the square roots of the eigenvalues $\mu_i$:
\be
{\cal A}^{l m}_{j k} \propto \sqrt{\mu_j \mu_k \mu_l \mu_m}\,.
\ee
This proportionality and the assumption that $\mu_1 \gg \mu_i$ allows to
order the elements of $\cal A$ in powers of the square roots of
$\mu_i$. Keeping only the leading order terms, i.e., ${\cal A}^{11}_{11}$ at zero order,
${\cal A}^{j1}_{11}$, ${\cal A}^{1j}_{11}$, ${\cal A}^{11}_{j1}$ and ${\cal
A}^{11}_{1j}$ at first order, and so on, we can approximate $\cal A$ by
\be
{\cal A}^{lm}_{jk}\simeq (T^{(\mathrm {qp})}_{lm})^* T^{(\mathrm {qp})}_{jk}\,,
\ee 
with 
\be
\label{Tqp}
T^{(\mathrm {qp})}_{jk}=\frac{{\cal A}^{11}_{jk}}{\sqrt{{\cal A}^{11}_{11}}}.
\ee
Thus the quasi-pure concurrence can be written as
\be
\label{conc_qp}
c(\rho)\simeq c_{\rm qp}(\rho)={\rm max} \left\{\sv_1-\sum_{i>1} \sv_i,0 \right\}\,,
\ee
where the $\sv_i$ are the singular values of the matrix $T^{(\mathrm {qp})}$ defined in
Eq.~(\ref{Tqp}).

With these tools at hand, entanglement in higher dimensional systems
can be explored in a computationally manageable way. We will use them
to monitor the entanglement dynamics under different sources of decoherence.

\section{Environment Models \label{sec_env}}

We start out with the assumption that each subsystem interacts only, and
independently, with its local environment and, therefore, any initially entangled state will evolve, in
some cases asymptotically, into a separable one. The dynamics of
this decoherence process can be described by the master equation
\begin{equation}
\label{eq:lindblad}
\frac{d{\rho}}{dt}=\left(
{\mathbbm 1}\otimes{\cal L}+{\cal L}\otimes{\mathbbm 1}\right)\rho,
\end{equation}
where $ \rho$ is the reduced density operator of the system. The interactions
of each component of the system with the environment are assumed to be
of the same form, represented by the Lindblad operator ${\cal L}$. Under the assumption of
complete positivity and Markovian dynamics, the action of $\cal L$ on
$\rho$ reads
\begin{equation}
\label{lindop}
{\cal L}  \rho=
\sum_i\frac{\Gamma_i}{2}\left(2\,L_i\,{\rho}\,L_i^\dagger - 
L_i^\dagger\,L_i\,{\rho} -
{\rho}\,L_i^\dagger\,L_i\right)\,.
\end{equation}
The operators $L_i$ and the rates $\Gamma_i$ define, respectively, the
specific type of the system-environment coupling, and its strength. Various physical
situations may arise when a system is coupled to a reservoir: dissipation can
take place, noise can be added to the system, or simply loss of phase
coherence can occur. All these processes correspond to a suitable
choice of the operators $L_i$ in Eq.~(\ref{lindop}) and lead to entanglement
decay, with a rate which is at the focus of this paper.

For two-level systems, the operators $L_i$ can be written in terms of the
Pauli matrices. In the case of an interaction with a thermal bath the
Lindblad operator reads
\begin{eqnarray}
\label{thermal}
{\cal L}\rho = \frac{\Gamma (\bar n +1)}{2} \left(2\, \sigma_- \rho \sigma_+ -
\sigma_+ \sigma_- \rho - \rho \sigma_+ \sigma_- \right) + \nonumber\\ \frac{\Gamma
  \bar n}{2}  \left(2\, \sigma_+ \rho \sigma_- -
\sigma_- \sigma_+ \rho - \rho \sigma_- \sigma_+ \right) \, .
\end{eqnarray}
In this equation, the first and the second term on the right hand side
describe, respectively, decay and excitation processes (mediated by
the two-level excitation and deexcitation operators $\sigma_+$ and $\sigma_-$), with rates which
depend on the temperature, here parametrized by $\bar n$, the average thermal
excitation of the reservoir. In the zero temperature limit, ${\bar n}=0$,
only the spontaneous decay term survives, leading to a purely dissipative
process, which corresponds to the fundamental limiting factor for the
coherent evolution of atomic qubits. Noisy dynamics corresponds to the infinite temperature limit, where
$\bar n\rightarrow\infty$, and, simultaneously, $\Gamma\rightarrow0$, so that
$\Gamma \bar n \equiv \tilde \Gamma$ is constant. In this case, decay and excitation
occur at exactly the same rate, and the noise induced by the transitions
between the two levels brings the system to a stationary, maximally
mixed state. Finally, a purely dephasing reservoir is obtained by choosing $L_i = d = \sigma_+
\sigma_-$, leading to the master equation 
\begin{equation}
\frac{d{\rho}}{dt}= \frac{\Gamma}{2} \left(2\, \sigma_+ \sigma_-
  \rho \sigma_+ \sigma_- - \sigma_+ \sigma_- \rho - \rho \sigma_+ \sigma_- \right) \, .
\label{deph}
\end{equation}
In this case, only the off-diagonal elements of the density matrix
decay, and phase coherence is lost. This dephasing mechanism is
known to be an important source of decoherence in ion
traps~\cite{monroe_95,blatt_04} and in quantum dots experiments~\cite{taylor_03,hayashi_03,itakura_03}.

If we consider a system of qudits, the operators $L_i$ in
Eq.~(\ref{lindop}) cannot be written anymore in terms of Pauli
matrices. To describe the effect of the environment in such systems,
we will consider the qudits as two bosonic modes with truncated bases
of length $d$. The
terms $n$ and $m$ in the general (pure) state
$\vert \psi \rangle =\sum_{n,m=0}^{d-1} \psi_{nm} \vert n\,m\rangle $
then indicate the
occupation number in each of these modes. Decay and excitation
processes now correspond to the action of annihilation and
creation operators $a$ and $a^{\dag}$, analogous to the $\sigma_-$ and
$\sigma_+$ for qubits. With this convention, the thermal and dephasing
environments are described, respectively, by
\begin{eqnarray}
\label{thermal2}
{\cal L}\rho = \frac{\Gamma (\bar n +1)}{2} \left(2\, a \rho a^{\dagger} -
a^{\dagger}a \rho - \rho a^{\dagger} a \right) + \nonumber\\ \frac{\Gamma
  \bar n}{2}  \left(2\, a^{\dag} \rho a -
a a^{\dagger} \rho - \rho a a^{\dagger} \right) \, ,
\end{eqnarray}
and 
\begin{eqnarray}
\label{deph2}
{\cal L}\rho = \frac{\Gamma}{2}  \left(2\, a^{\dag} a \rho a^{\dag} a -
a^{\dagger} a \rho - \rho a^{\dagger} a  \right) \,.
\end{eqnarray}

Zero and infinite temperature limits are obtained as in the qubit
case, and represent, to a very good approximation, the decoherence
processes of photons in high-$Q$ cavities~\cite{brune_96}, and of motional degrees of freedom in ion traps~\cite{turch00}.

\section{Entanglement dynamics under decoherence \label{sec_dyn}}

\subsection{Two qubits case \label{sec_qubits}}

We shall begin our analysis with the case of two qubits, since it
allows for analytical solutions, and can provide some intuition of the
processes at work in higher dimensional systems. To provide a more complete and
systematic description of the timescales of entanglement decay under
environment interaction, we will recall some known results for the case of
dephasing and zero temperature reservoirs, and furthermore derive exact solutions when
finite temperature effects are taken into  account.

Let us start with initially prepared Bell states $\vert \Psi^{\pm}\rangle=\left(\vert 01 \rangle \pm \vert 10
  \rangle\right)/\sqrt{2}$, and $\vert \Phi^{\pm}\rangle=\left(\vert 00 \rangle \pm \vert 11
  \rangle\right)/\sqrt{2}$. Their concurrencies
  as a function of time are given by
\begin{equation}
\label{conc_deph}
c(\Psi^{\pm},t)=c(\Phi^{\pm},t)=e^{-\Gamma t},
\end{equation}
for the dephasing environment, and by
\begin{equation}
\label{conc_T0}
c(\Psi^{\pm},t)=e^{-\Gamma t}, \:\:\: c(\Phi^{\pm},t)=e^{-2 \Gamma t}\,,
\end{equation}
for the zero temperature case. 
The first important observation is the
accelerated (by a factor of two) decay of concurrence for the $\Phi^{\pm}$ as
compared to the $\Psi^{\pm}$ states, under the influence of a zero temperature
environment. This can be understood from the timescales involved in the
corresponding solution for the density matrix: while for $\Psi^{\pm}$ each term
$\ket{01}$ and $\ket{10}$ corresponds to a single particle decay, leading to a timescale $e^{-\Gamma t}$, we have the
term $\ket{11}$ in $\Phi^{\pm}$, such that both particles can undergo an
environment induced transition to the ground state, thus introducing a faster,
$e^{-2 \Gamma t}$, decay.

Also for the finite temperature case can an explicit solution be
calculated for initial Bell states: it reads
\begin{equation}
\label{conc_T}
c(t)={\rm max}\left\{c_T(t),0\right\}\,,  
\end{equation}
where the function $c_T(t)$, for $\Psi^{\pm}$ and $\Phi^{\pm}$ states, is
given, respectively, by
\begin{equation}
\label{conc_T_psi}
c_T(\Psi^{\pm},t)=\beta-\frac{2(1-\beta)\sqrt{(\nb^2+\nb)^2\left(\beta+1\right)^2+\beta(\nb^2+\nb)}}{(2\nb+1)^2}\,,
\end{equation}
and 
\begin{equation}
\label{conc_T_phi}
c_T(\Phi^{\pm}t)=\beta+\frac{\left[2\nb(\nb+1)+1\right]\beta^2-\beta-2\nb(\nb+1)}{(2\nb+1)^2}\,,
\end{equation}
with $\beta=e^{-\Gamma(2\nb+1) t}$. These expressions, which reveal the
precise role of temperature on the decay of entanglement of Bell states, show
that, distinct from the cases of dephasing and zero temperature
environments, concurrence does not decay according to a simple
mono-exponential law, but rather exhibits various timescales. To apprehend the essential features of temperature
effects on entanglement decay, it is useful to focus on the short time
limit of these equations~\cite{retamal_04}. Expanding Eqs.~(\ref{conc_T_psi}) and (\ref{conc_T_phi}) to first order in $t$, one
finds that the short time concurrence decay is given by 
\begin{equation}
c(\Psi^{\pm},t)\simeq 1-\left(2\bar n+1 +2\sqrt{\bar n (\bar n
    +1)}\right)\Gamma t , 
  \end{equation}
and
  \begin{equation}
    c(\Phi^{\pm},t)\simeq 1-2(2\bar n+1) \Gamma t \,. 
  \end{equation}
 This shows that, at the beginning of the evolution, a finite temperature
    reservoir increases the concurrence decay rate, as compared to the zero
    temperature case, by a factor of $\left(2\bar n+1 +2\sqrt{\bar n (\bar n
    +1)}\right)$ and of $(2\bar n+1)$, for $\Psi^{\pm}$ and $\Phi^{\pm}$
    states, respectively.

From Eqs.~(\ref{conc_T_psi}) and
(\ref{conc_T_phi}) one can also derive the infinite temperature limit ($\bar n\rightarrow\infty$, $\Gamma\rightarrow0$, with $\Gamma \bar n \equiv \tilde
\Gamma=\textrm {const.}$) of the thermal bath. In this case, the concurrence becomes
\begin{equation}
\label{conc_TI}
c(\Psi^{\pm},t)=c(\Phi^{\pm},t)={\rm max}\left\{\frac{e^{-4\tilde \Gamma t}}{2}+e^{-2\tilde \Gamma
t}-\frac{1}{2},0\right\}\,.  
\end{equation}
Note that, again, entanglement dynamics involves different
timescales. However, a single exponential captures the basic behavior
of entanglement decay, since Eq.~(\ref{conc_TI}) can be well fitted by
a function in the form $\alpha e^{-\gamma t}+ \delta$, with negative offset
$\delta$. Consequently, separability, i.e. $c(t)=0$, is reached at finite times
for the infinite temperature environment, as well as for any $\bar n >0$. This is in contrast to the
above dephasing and zero temperature environments, where separability
is reached only asymptotically, and follows from the long-time limit for $c_T(t)$ in
Eq.~(\ref{conc_T}) 
\begin{equation}
\lim_{t \to \infty} c_T(t) = -\frac{2\bar
      n(\bar n+1)}{(2\bar n+1)^2}\,,
\end{equation}
which is non-positive for all $\bar n>0$. The exact expressions for this separability
time can be easily obtained from the condition $c(t)=0$, and are of the form
$t_{\rm sep}=\ln{\left[f(\nb)\right]}/\Gamma(2\nb+1)$, where the
function $f(\nb)$ depends on the initial state $\Psi^{\pm}$ or
$\Phi^{\pm}$. Observe, however, that this is not true for arbitrary
initial states. Indeed, some initially mixed states~\cite{eberly_04} as well
as some pure non-maximally entangled states, e.g., $\vert \psi
\rangle=\frac{1}{2}\vert 00\rangle +\frac{1}{2}\vert
01\rangle +\frac{1}{\sqrt{2}}\vert 11\rangle$, reach separability on
finite timescales even for a zero temperature environment.   

\subsection{Two qudits case \label{sec_qudits}}

As the dimensions of the subsystems increase, not only a general
analytical solution is unknown, but also the numerical approach to
obtain reliable estimates of concurrence rapidly turns into a very demanding
task. However, it is exactly in this situation where the strength of
the tools derived in Section~\ref{sec_conc}, for calculating lower
bounds of concurrence, becomes manifest. We will devote the
remainder of this paper to a systematic analysis of entanglement
dynamics in bipartite qudits, using these tools. In
particular, we will extract bounds for the {\em decay rates} of entanglement,
and also show how one can infer analytical solutions even for arbitrary
dimensions, from the knowledge of the dynamics of the
system.

\subsubsection{Dephasing environment: an exactly solvable example}

We begin our analysis with the dephasing dynamics described by
Eq.~(\ref{deph2}), which, as discussed in Section~\ref{sec_env},
induces a decay of the off-diagonal elements of the density matrix
without changing the diagonal ones. Besides its importance in the
description of different physical situations, as already mentioned
before, the dephasing model is very instructive since it allows for obtaining analytical solutions for the concurrence of
some classes of states. Let us first assume that the initial pure state is of the form $\vert \psi\rangle= a \vert m_1\,m_2\rangle+ b \vert
n_1\,n_2\rangle$, what comprises two-level Bell states as special
cases. In this situation, one can write the solution of Eq.~(\ref{deph2}) as
\bqa
\label{rhodeph}
\rho(t)=\vert a\vert^2 \ket{m_1m_2}\bra{m_1m_2}+ \vert b\vert^2
\ket{n_1n_2}\bra{n_1n_2} \nonumber \\ + e^{-\gamma t} ab^*\ket{m_1
m_2}\bra{n_1 n_2} + e^{-\gamma t} a^*b
\ket{n_1 n_2}\bra{m_1 m_2},
\eqa
with 
\be
\gamma=\frac{\Gamma}{2}\left[(m_1-n_1)^2+(m_2-n_2)^2\right].
\ee

Note that there is just one decaying element (and its conjugate) in the
density matrix evolution and, therefore, one expects that the
entanglement decay rates should follow the same behavior. This is indeed the
case, as we will show using the techniques described in
Section~\ref{sec_conc}: First, one can easily check that the mixed state
(\ref{rhodeph}) can be decomposed into contributions of two pure states,
\bqa
\ket{\phi_1}=\sqrt{p}\left( a \ket{m_1m_2}
+ b \ket{n_1n_2}\right)\, ,
\eqa
and
\bqa
\ket{\phi_2}=\sqrt{1-p}\bigl( a \ket{m_1m_2}
- b \ket{n_1n_2}\bigr)\, ,
\eqa
with $p=(1+e^{-\gamma t})/2$, and $\rho= \ket{\phi_1}\bra{\phi_1}+\ket{\phi_2}\bra{\phi_2}$. From the structure of this decomposition, one
can see that, among all $\ket{\chi_{\alpha}}$,
only 
$\ket{\chi}=(\ket{m_1 n_1} -\ket{n_1
  m_1})\otimes(\ket{m_2 n_2} -\ket{n_2 m_2})$ gives a non-zero
contribution for  $T^{\alpha}$ in Eq.~(\ref{talpha}). In this
case, the concurrence can be expressed in terms of a single matrix $T$,
which reads 
\begin{displaymath}
T=\left(
\begin{array}{cc}
2 p\, a\, b & 0\\
0 & 2 (p-1)\, a\, b
\end{array}
\right).
\end{displaymath}
Thus the formal structure of two-level
systems is retrieved, and the lower bound, Eqs.~(\ref{lbound}) and (\ref{freeparbound}), is exact. From the singular values of the matrix $T$, one readily obtains
\be
\label{dephexact}
c(t)= 2 \vert a b\vert \, e^{\left(-\Gamma
t/2\right)\left[(m_1-n_1)^2+(m_2-n_2)^2\right]}\,,
\ee
which is equal to the sum of the absolute values of the two non-vanishing
off-diagonal elements of $\rho(t)$. 

\subsubsection{Zero temperature environment: exact solutions, and estimates}

Although there are no a priori criteria
from which to infer the existence of exact solutions like Eq.~(\ref{dephexact})
in general, this is still possible for a large variety of physical
situations. For a zero temperature environment, for example, exact solutions
can be derived based on a given form of the decomposition of $\rho$. Moreover, even if such decomposition does not exist, it is still
possible to obtain estimates on the decay rates of concurrence with the tools
described in Section~\ref{sec_conc}.

To illustrate this, we start with states of the form $\vert \psi\rangle= a
\vert 0 m\rangle+ b\vert m 0\rangle$ and consider, for the derivation
of the exact solution, the case of three-level systems and $a=b=1/\sqrt{2}$. The
decomposition of $\rho(t)$ in terms of its eigenstates (with non-vanishing
eigenvalues) reads
\bqa
\ket{\phi_1}=e^{-\Gamma t} \left(\ket{0 2}
+ \ket{20}\right)/\sqrt{2}\, ,
\eqa
\bqa
\ket{\phi_2}=\sqrt{e^{-\Gamma t}\left(1-e^{-\Gamma t}\right)} \ket{01}\, ,
\eqa
\bqa
\ket{\phi_3}=\sqrt{e^{-\Gamma t}\left(1-e^{-\Gamma t}\right)} \ket{10}\, ,
\eqa
and
\bqa
\ket{\phi_4}=\left(1-e^{-\Gamma t}\right) \ket{00}\, .
\eqa
All these states, with the exception of $\ket{\phi_1}$, are
separable. Therefore, from Eq.~(\ref{mixwotcon}), one can see that
the concurrence of $\ket{\phi_1}$ is an upper bound of
$c(\rho)$. Moreover, this term is also the only one that gives a non-vanishing contribution to the matrix $T^{(\mathrm {qp})}$, Eq.~(\ref{Tqp}),
in the quasi-pure approximation. Hence, lower and upper bounds
coincide and the solution $c(t)=e^{- 2 \Gamma t}$ is exact. For the
state $\vert \psi\rangle= a \vert 0 m\rangle+ b\vert m 0\rangle$ the demonstration is analogous, and the concurrence is given by
\be
\label{conc_0m_m0}
c(t)=2\vert a b \vert \, e^{- m \Gamma t}.
\ee
From the time-dependent matrix elements of $\rho$, which can be obtained from
Eq.~(\ref{thermal2}) with $\bar n=0$, one can see that, as in Eq.~(\ref{dephexact}), concurrence is
simply given by the sum of the absolute values of the only two non-vanishing off-diagonal elements of $\rho$. However, we should
emphasize that this only happens in special cases alike the present
one and, in general, there is no
simple relation between the decay of off-diagonal elements of $\rho$ and
concurrence. 

An example where such a simple relation does not exist, even though an
exact solution is possible, is the zero temperature dynamics of
initial states of the form $\vert \psi\rangle= a \vert 0 0\rangle+ b \vert m
m\rangle$. 
The time-dependent density matrix can be decomposed as $\rho= \xi +
\eta$, with $\eta$ diagonal in the $\{\vert i\,j\rangle\}$ basis (and
thus separable), and $\xi=\sum\matel{ij}{\rho}{pq}\ket{ij}\bra{pq}$
with $i,\, j,\,p,\,q \in \{0,m\}$. From this argument and from
Eq.(\ref{mixwotcon}) it follows that the concurrence of $\xi$ is an upper
bound of $c(\rho)$. Furthermore, $\xi$ is a two-qubit-like matrix and its
concurrence can be readily obtained as
\begin{eqnarray}
\label{conc_00mm}
c(t)=2 e^{-m\Gamma t}\left(\vert a b\vert -\left(1-e^{-\Gamma
      t}\right)^m \vert b\vert^2\right)\,.
\end{eqnarray}
This result coincides with the quasi-pure approximation of
$c(\xi)$, calculated using
$\ket{\chi}=(\ket{0m}-\ket{m0})\otimes(\ket{0m}-\ket{m0})$ in Eqs.(\ref{talpha}), (\ref{tensor_A}),  (\ref{conc_qp}), and hence is exact.

It is clear from Eq.~(\ref{conc_00mm}) that the entanglement decay, in
this case, encompasses different timescales for $d>2$, while the only
non-zero off-diagonal elements, $\bra{00}\rho\ket{mm}$ and
$\bra{mm}\rho\ket{00}$, decay as $e^{-m\Gamma t}$. 
Note also that, in the limit of large $m$, the concurrence tends to
$c(t)=2 \vert a b \vert e^{-m \Gamma \, t}$ and then coincides with
the result of Eq.~(\ref{conc_0m_m0}). Thus, in this limit, the state $\vert
\psi\rangle= a \vert 0 0\rangle+ b \vert m m\rangle$
decays as the state $\vert \psi\rangle= a \vert 0 m\rangle+ b \vert m
 0\rangle$. This shows that the robustness of
 the $\Psi^{\pm}$ Bell states as compared to the $\Phi^{\pm}$ in terms of
 decay rates in the two qubit case (see Eq.~(\ref{conc_T0})) fades away for
 larger dimensions. 

In the two previous examples, exact solutions were possible because $\rho$
could be decomposed into a separable part, and a non-separable two-level-like
one whose upper and lower bounds coincide. Although, at first sight, one might
judge these conditions to be rather restrictive, the above considerations can
be also applied to other environment models and qubit-like initial
states. However, what happens when such decomposition does not exist? How much
information on the decay rates can we extract with the available tools?

\begin{figure}
\includegraphics[width=8.0cm]{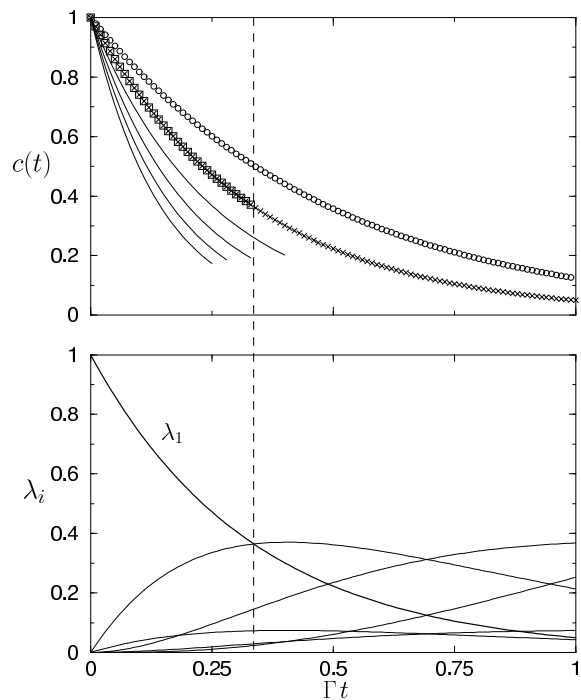}
\caption{Top panel: optimized upper (circles) and lower (crosses) bounds of
concurrence, as well as quasi-pure approximation (squares), for the initial
state $\ket{\psi}=\left(\ket{12}+\ket{21}\right)/\sqrt{2}$ coupled to a zero temperature reservoir. The solid lines show the quasi-pure
approximations for initial states
$\ket{\psi}=\left(\ket{1m}+\ket{m1}\right)/\sqrt{2}$, with $m=d-1$
and, from top to bottom, $d=4$ to $d=7$. Quasi-pure solutions behave
as $e^{-(m+1)\Gamma\, t}$ and the upper bound for $d=3$ is correctly fitted by $e^{-2\Gamma\, t}$.
Bottom panel: eigenvalues of the density matrix
$\rho(t)$ as a function of time for $d=3$. Quasi-pure results
coincide with the optimized lower bound and are only calculated while
$\lambda_1$ remains the largest eigenvalue.} 
\label{fig1}
\end{figure}

To help to answer these questions let us focus on the situation shown in Fig.~\ref{fig1}, where an
 analytical solution could not be found. For the initial state
 $\ket{\psi}=\left(\ket{12}+\ket{21}\right)/\sqrt{2}$, upper~(\ref{mixwotcon}) and
 optimized lower bounds~(\ref{lbound}, \ref{freeparbound}), as well as
 quasi-pure calculations~(\ref{conc_qp}), are presented in the upper panel of the figure, as
 circles, crosses, and squares, respectively. The solid lines
 represent the quasi-pure approximations for initial states of the form
 $\ket{\psi}=\left(\ket{1m}+\ket{m1}\right)/\sqrt{2}$, with $m=d-1$
 and, from top to bottom, $d=4$ to $d=7$. Note that the quasi-pure
 approximations are calculated only until times when the initially
 largest eigenvalue $\lambda_1$ coincides with the second one (see the
 lower
 panel of Fig.~\ref{fig1} for the case $d=3$). The quasi-pure solutions solutions have a simple form
$e^{-(m+1)\Gamma\,t}$. For $d=3$, we can go further and use also the
numerically calculated upper bound, which in this case is found (through
curve fitting) to behave as $e^{-2\Gamma t}$, to confine the actual
value of the decay rate to the interval $2 \Gamma \le \gamma \le 3
\Gamma$. Furthermore, we can state that for arbitrary dimensions the
decay rate cannot exceed the value given by quasi-pure approximation,
which, in the present case, leads to $\gamma \le d \Gamma$. This is a noticeable asset of our analysis.
Even if the high dimension of the system, together with the complex structure of the states, do not allow
to determine the precise decay rates of entanglement in the
general case, we can extract useful information on their bounds.

\subsubsection{Finite temperature environment}

Finally, let us briefly present a situation where the ability to handle higher
dimensional systems is essential: the finite temperature reservoir. Any
initial state will evolve asymptotically into a thermal state, which can be
described by a finite number of levels -- which, however, increases with temperature.
 
The decoherence effect of this environment on qudits is illustrated in Figure~\ref{fig2}, for an initial state
$\ket{\psi}=(\ket{01} + \ket{10})/\sqrt{2}$. The solid, dashed and long-dashed
lines represent, respectively, the two qubit solution,
Eq.~({\ref{conc_T_psi}), for $\nb=0.1$, $\nb=0.2$, and the infinite temperature
  limit. The corresponding quasi-pure solutions for the same initial state
  evolving under Eq.~(\ref{thermal2}) are given by circles, squares and
  crosses. The bold solid curve indicates, for comparison, the zero
  temperature solution. The dimension used in the simulations ($d=8$) was
  chosen to be large enough to describe the dynamics of the system for $\bar
  n=0.1$ and $\bar n=0.2$, for the times shown in the figure. Note that for the
  infinite temperature case actually an infinite basis is needed and
  that our truncated basis set allows for a correct description of the
  dynamics only for a short period of time (${\tilde \Gamma} \,t \approx
 0.06$), as long as the dynamics do not populate levels at the
 boundary of the Hilbert space.

The most noticeable effect here is that the expected enhancement of
 the decay rates
 with temperature is more pronounced for qudits than for qubits. As a matter of
 fact, this is not too surprising since the dynamics of a finite temperature bath,
 given by Eq.~(\ref{thermal2}), induces transitions to many different levels,
 thus speeding up entanglement decay.

\begin{figure}
\includegraphics[width=8.0cm]{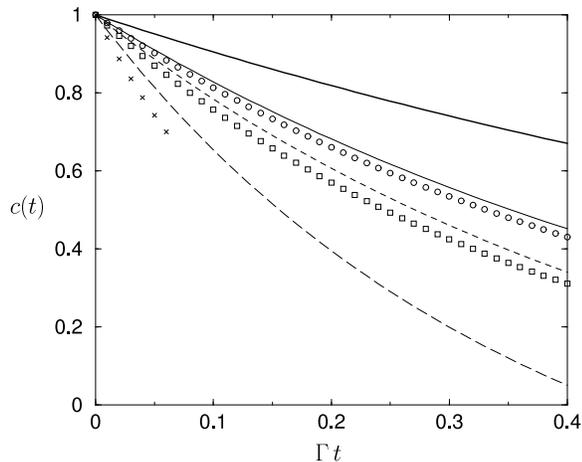}
\caption{Finite temperature effects on the decay of entanglement for an initial
Bell state $\Psi^{+}$. Solid, dashed and long dashed lines are, respectively, the two-level
case solutions of Eq.~(\ref{thermal}), given by Eq.~(\ref{conc_T_psi}), for
$\nb=0.1$, $\nb=0.2$, and the infinite temperature limit. Circles, squares and
crosses are the corresponding quasi-pure results for the qudits' dynamics given
by Eq.~(\ref{thermal2}) using a truncated basis with $d=8$. For
comparison, also the solution for a zero temperature environment is
shown (bold line). The expected enhancement of the decay rates with
increasing temperature is more pronounced in the higher dimensional
case.}
\label{fig2}
\end{figure}

\section{Conclusions}
\label{sec_concl}

In this paper we have contributed to the understanding of the
dynamics of entanglement under environment coupling for systems of two qudits. 
Although no analytical solutions are available in general, we have
shown that they can be derived for different classes of initial states and dynamics. 
Moreover, we have
shown how suitable decompositions of the density
matrix, together with techniques for calculating lower bounds of
concurrence, can be useful to decide on the existence of exact
solutions and to derive them for concurrence or its lower bounds. For
more general cases in high dimensional
systems, we were able to derive upper bounds for the decay rates of entanglement.

Future work will have to focus on error estimates for the employed
approximations for concurrence, in higher dimensions.

\bibliography{literaturaQI}

\end{document}